\documentclass[fleqn,12pt]{wlscirep}
\usepackage{amsmath}
\usepackage{graphicx}
\usepackage{footmisc}
\usepackage{color}

\begin{document}
\title {The ground state of the  Kondo lattice } 
\author[1]{Igor N. Karnaukhov} 
\affil[1]{G.V. Kurdyumov Institute for Metal Physics, 36 Vernadsky Boulevard, 03142 Kyiv, Ukraine}
\affil[*]{karnaui@yahoo.com}
\begin{abstract}
The Kondo insulator state (KIS)  is among the least understood phase state in condensed matter physics.
KIS is the state of the electron liquid in the Kondo lattice at half filling, is studied within the mean field approach. 
We demonstrate, that $Z_2$-field, which is formed  by electrons and local moments, leads to the state of the Kondo insulator in a lattice with a double cell. In the ground state, electrons and local moments form singlets; in this case, no spin or charge density waves are realized in a lattice with  a double cell.  We have shown that a Majorana-type gap spectrum of  the quasi-particle excitations is realized in $Z_2$-field.
The gap in the spectrum decreases with increasing external magnetic field; it closes at a critical value at the insulator-metal phase transition point.  Thus, the introduction of the $Z_2$ -field allows you to answer the key question what is the ground state of  KIS.  
\end{abstract}
\maketitle

\section*{Introduction}

In contrast to the Kondo problem, the exact solution of which was obtained by  a weak interaction in continuum approach \cite{1,2},  the behavior (phase state) of an electron liquid in the Kondo lattice  is an unsolved problem of condensed matter physics.  The scattering electrons on a  local  monent with spin flip leads to the Abrikosov-Suhl resonance, a new behavior of an  electron liquid in the Kondo problem  at low temperatures and magnetic field.  

Speaking of the Kondo lattice, we do not know the answers to simple but important questions:  what is the ground state of the electron liquid in KIS, why is there a large Fermi surface at conservation  of the number of electrons, what is the nature of the charge and spin gaps in the excitation spectrum \cite{CA,3,K1,K2,4,5,K3}. 
When solving the Kondo lattice problem, it is also necessary to take into account the  scattering of electrons  by local moments with spin  flip (as it takes place in the Kondo problem). The effective Hamiltonian should also not break the symmetry of the model. This is a non-trivial problem that has not yet been solved, so we cannot say anything definitive about what the ground state is implemented in KIS.

However, despite this pessimistic introduction, the purpose of the article is to answer a main question, what is the ground state of the Kondo lattice at half filling.
The antiferromagnetic exchange interaction between electrons and local moments leads to a $Z_2$-field whose uniform configuration forms a lattice with a double cell  in  KIS. Note that charge or spin density waves are not realized in a lattice with a double cell; in this sense, this is an unusual phase state.

\section*{Model}
The Hamiltonian of the spin-$\frac{1}{2}$ Kondo lattice dimension D ${\cal H}={\cal H}_0+{\cal H}_{K}$ includes two terms, the first of which is determined by energy of electrons,  the second is determined by the contact exchange interaction of these electrons with moments located at the lattice sites 
\begin{eqnarray} &&{\cal H}_0= - \sum_{<i,j>}\sum_{\sigma=\uparrow,\downarrow}c^\dagger_{i \sigma} c_{j \sigma}-2h \sum_{j}(s^z_j+S^z_j)
, \nonumber \\ 
&&{\cal H}_K= 2\sum_{j=1}^N [J s^z_jS^z_j+K (s^x_jS^x_j+s^y_jS^y_j)], \end{eqnarray} 

where $c^\dagger_{j \sigma}$ and $c_{j \sigma}$ are the fermion operators determined on a lattice site $j$, $\sigma =\uparrow,\downarrow$ denotes the spin of electron, the hopping integral between the nearest-neighbor lattice sites is equal to one, the spin operators of electrons $s_j^{\alpha}=\frac{1}{2}c^\dagger_{j \sigma}\sigma^{\alpha}_{\sigma \sigma'}c_{j\sigma'}$ are determined by the Pauli matrices $\sigma^{\alpha}$ ($\alpha=x,y,z$), $\textbf{S}_j$ is the spin-$\frac{1}{2}$ operator defined on the lattice site $j$ ($S$ is its value), $J\geq 0$ and $K>0$ are the magnitudes of the exchange interaction ($K=J>0$ corresponds to an isotropic antiferromagnetic exchange  interaction,  $J=0, K>0$ corresponds to a strong anisotropic interaction), $h$ is an external magnetic field ($g-$ factor is 2, we assume the Bohr magneton is 1), N is the total number of lattice sites.  We study the behavior of an electron liquid in the chain (1D) and on the square (2D) and cubic (3D) lattices at half-filling.

\section*{The ground-state of the Kondo lattice}

We consider the term of the Hamiltonian (1) ${\cal H}_{K}$ in detail in the following form \begin{eqnarray} &&{\cal H}_K= \sum_{j=1}^N[2J s^z_jS^z_j+K(s^+_jS^-_j + s^-_jS^+_j)], \end{eqnarray} 
where the spin operators redefine via the fermion operators, $s^z_j=\frac{1}{2}(m_{j\uparrow}-m_{j\downarrow})$,   $s^+_j=c^\dagger_{j \uparrow} c_{j\downarrow}$, $s^-_j=c^\dagger_{j \downarrow} c_{j\uparrow}$,  here $m_{j\sigma}=c^\dagger_{j \sigma} c_{j \sigma}$, $m_{j}=m_{j\uparrow}+m_{j\downarrow}$ are the density operators.

We use the  following presentation for the ${\cal H}_{K}$ term:
$2Js^z_jS^z_j+K(s^+_jS^-_j + s^-_jS^+_j)=-J (s^z_j-S^z_j)^2-K(s^+_j-S^+_j)^\dagger(s^+_j-S^+_j) +(J-K) (S^z_j)^2-\frac{1}{4}(J+2 K)m_{j}^2+\frac{1}{2}(J+K)m_j+K S(S+1)
 \Longrightarrow 2\Lambda_{j} (s^z_j-S^z_j)+\lambda_{j}(c^+_{j\uparrow}c_{j\downarrow}-S^+_{j})+\lambda^*_{j}(c^+_{j\downarrow}c_{j\uparrow}- S^-_{j} )+2\mu_{j} m_{j}$. Using  the Hubbard-Stratanovich  transformation we introduce  the effective Hamiltonian which is determined by two component $\Lambda-\lambda$-field. We will study in detail the case $S=\frac{1}{2}$.
  $ (S^z_j)^2$ -operator is conserved and  $\mu$-component shifts the Fermi energy, so they can be neglected.  We can define an effective Hamiltonian ${\cal H}_{eff}$, which describes the behavior of the electron liquid  in the Kondo lattice in the mean field approach

${\cal H}_{eff}={\cal H}_0
\sum_{j}[2\Lambda_{j} (s^z_j-S^z_j)+\lambda_{j}(c^+_{j\uparrow}c_{j\downarrow}-S^+_{j})+\lambda^*_{j}(c^+_{j\downarrow}c_{j\uparrow}-S^-_{j}) +\sum_j[\frac{\Lambda_j^2}{J}+ \frac{|\lambda_{j}|^2}{K}]$.

Let us consider equations for the one-particle wave functions $\psi(\textbf{j},\sigma)c^\dagger_{\textbf{j} \sigma}+\phi (\textbf{j},\pm\sigma)S^{\pm}_{\textbf{j} }$  ($\sigma=\uparrow,\downarrow)$ with energy $\epsilon$, the  $\psi(\textbf{j},\sigma)$ and $ \phi(\textbf{j},\sigma)$  amplitudes satisfy the following equations : 
\begin{eqnarray} 
&&(\epsilon -\Lambda_{\textbf{j}})\psi(\textbf{j},\sigma) +\lambda_{\textbf{j}} \psi(\textbf{j},-\sigma)+\sum_{\textbf{1}}\psi(\textbf{j+1},\sigma)=0, \nonumber\\
 &&(\epsilon + \Lambda_{\textbf{j}})\psi(\textbf{j},-\sigma +\lambda^*_{\textbf{j}}\psi(\textbf{j},\sigma)+ \sum_{\textbf{1}}\psi(\textbf{j+1},-\sigma)=0, \nonumber\\ 
&&(\epsilon +\Lambda_{\textbf{j}})\phi(\textbf{j},\sigma) -\lambda_{\textbf{j}} \phi(\textbf{j},-\sigma)=0,\nonumber\\ 
&&(\epsilon -\Lambda_{\textbf{j}})\phi(\textbf{j},-\sigma) -\lambda^*_{\textbf{j}} \phi(\textbf{j},\sigma)=0, 
\end{eqnarray}

 where sums over the nearest lattice sites. The variables $\Lambda_{\textbf{j}}=\pm \Lambda$ and $\lambda_{\textbf{j}}=\pm \lambda$ are identified with a static two component $\mathbb{Z}_2$- field determined on the lattice sites. A  confuguration of this field, which corresponds to  an energy minimum, defines the ground state. The local moments form the flat band states, with the energies $\varepsilon_S=\pm\sqrt{\Lambda^2+\lambda^2}$.  The moments are located at the lattice sites, their energy is not determined by its spin.

Detailed numerical analysis shows, that an uniform sector with $\Lambda_\textbf{j} =\Lambda$, $\lambda_\textbf{j} =\lambda$ and $\Lambda_\textbf{j+1}=-\Lambda$, $\lambda_\textbf{j+1}=-\lambda$  corresponds to the ground state of an electron liquid for arbitrary values of $J$  and $K $ \cite{K4,K5,Lieb,Kitaev}.
Using Eqs(3) we calculate the energy of quasi-particle excitations wich correspond to this uniform configuration of the $Z_2$-field,. The spectrum includes two branches of local moments $\varepsilon_S$ and two branches of electrons $\varepsilon_s(\textbf{k})=\pm\sqrt{\Lambda^2+\lambda^2+|w(\textbf{k})|^2}$, 
here $w(\textbf{k})=\sum_{\alpha}^D[1+\exp ( i k_\alpha)]$, $\textbf{k}=(k_x,k_y,k_z)$ is the wave vector.
The solution for the values of the $Z_2$-field components satisfies the energy minimum or the saddle point of the action. For isotropic ($J=K>0$,  here $\Lambda=\lambda\neq 0$) and anisotropic ($J=0, K>0$, here $\Lambda=0, \lambda \neq 0$) exchange interaction, a self-consistent equation has the following form at $T=0K$ 

 \begin{equation} \frac{2\lambda}{J}=\frac{1}{N}\sum_{\textbf{k}}\frac{\lambda}{|\varepsilon_s(\textbf{k})|}+\frac{\lambda}{|\varepsilon_S|}. \label{eq:H6} \end{equation}

The behavior of the electron liquid in the case of strongly anisotropic, when $J=0$, $K>0$, and isotropic, when $J=K>0$, exchange  antiferromagnetic interaction will be considered in detail.

\begin{figure}[tp] \centering{\leavevmode} \begin{minipage}[h]{.45\linewidth} \center{ \includegraphics[width=\linewidth]{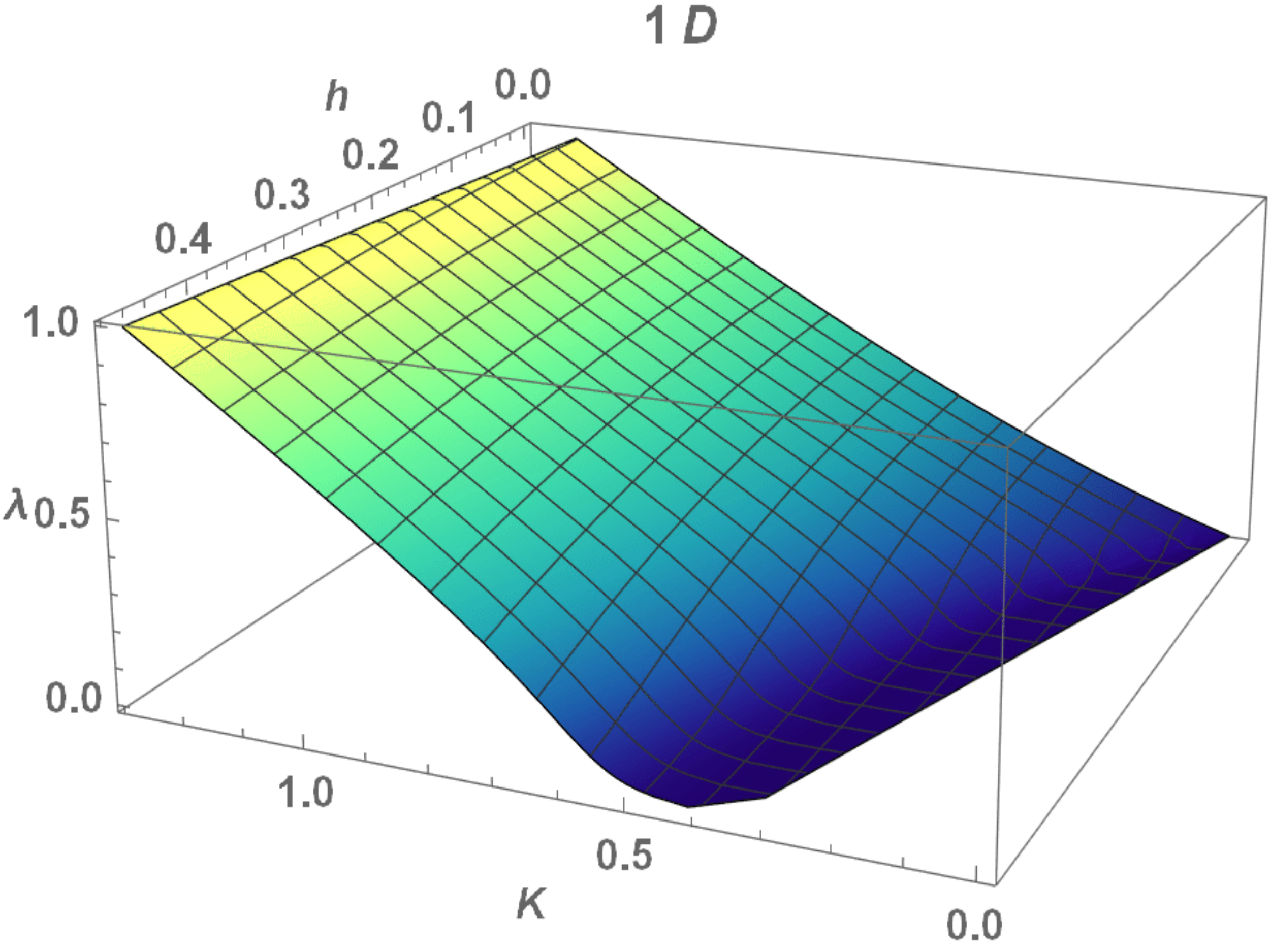} a)\\ 
} \end{minipage} \begin{minipage}[h]{.45\linewidth}\center{ \includegraphics[width=\linewidth]{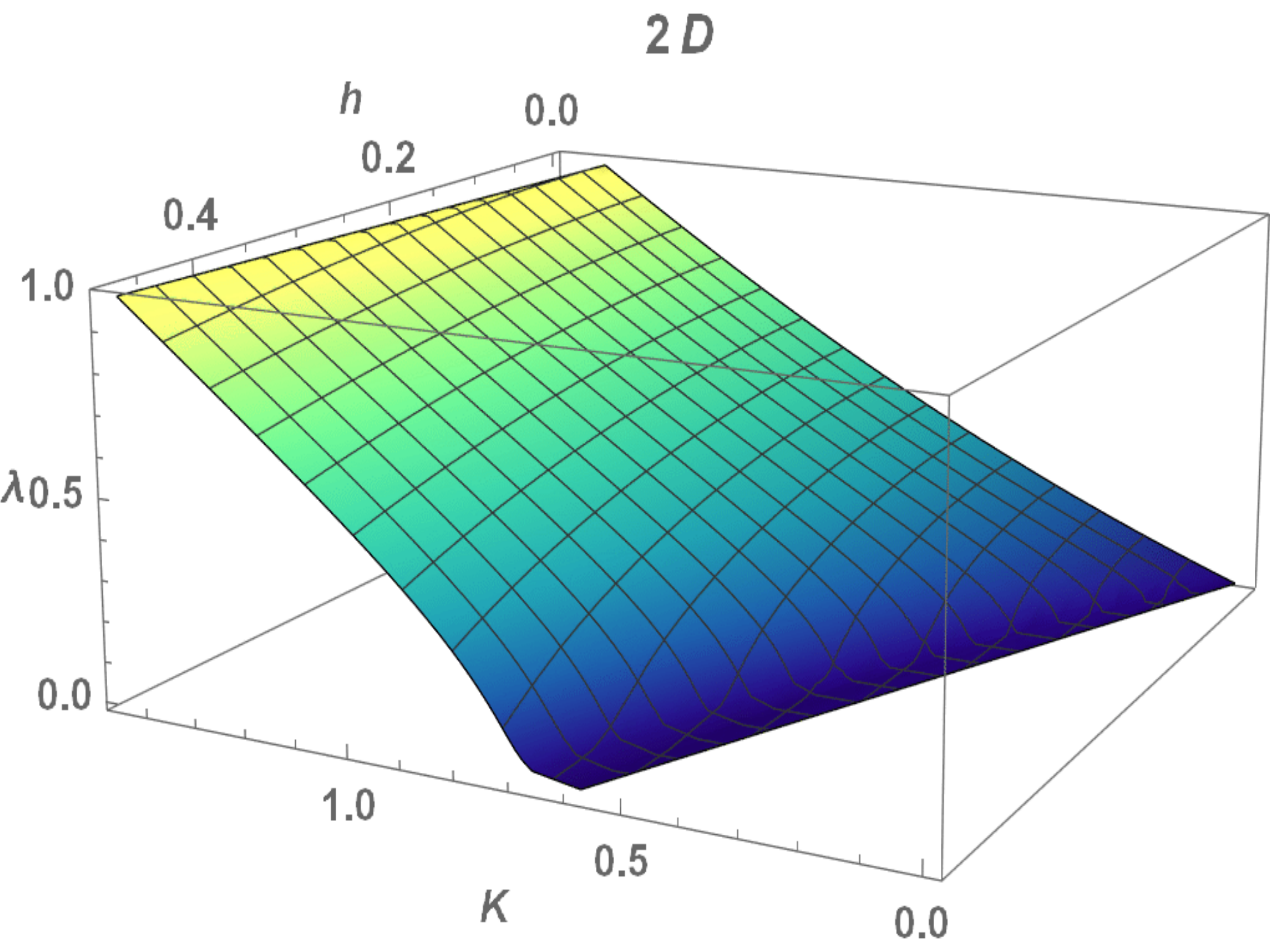} b)\\ 
} \end{minipage} \begin{minipage}[h]{.5\linewidth}
 \center{ \includegraphics[width=\linewidth]{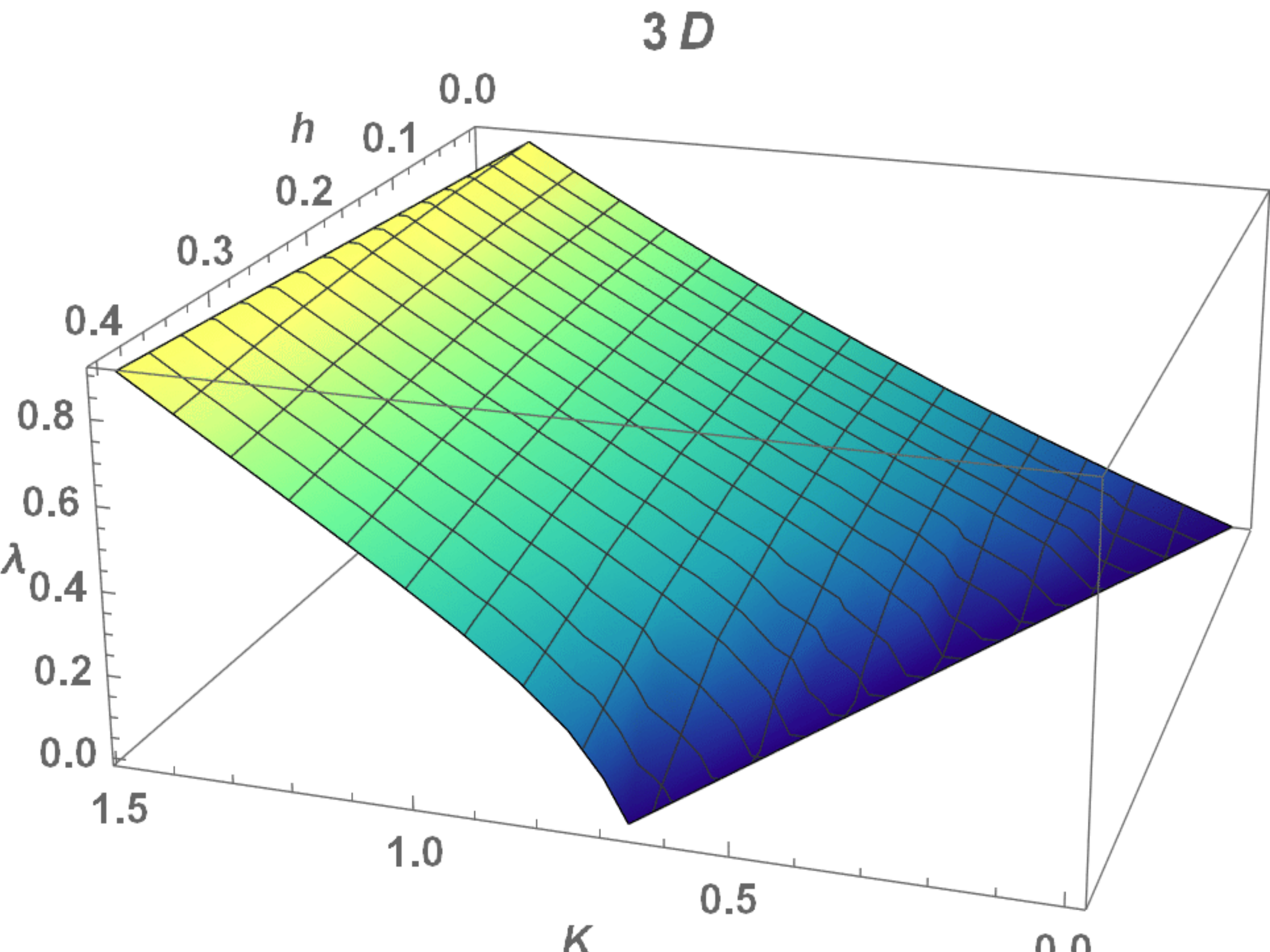} c)\\ 
} \end{minipage} \caption{(Color online)  $\lambda-$value as a function of   the exchange  integral $K$ and magnetic field $h$ , calculated for the  chain a),  square b) and cubic c)  lattices . } \label{fig:1} \end{figure}

\begin{figure}[tp] \centering{\leavevmode} \begin{minipage}[h]{.47\linewidth} \center{ \includegraphics[width=\linewidth]{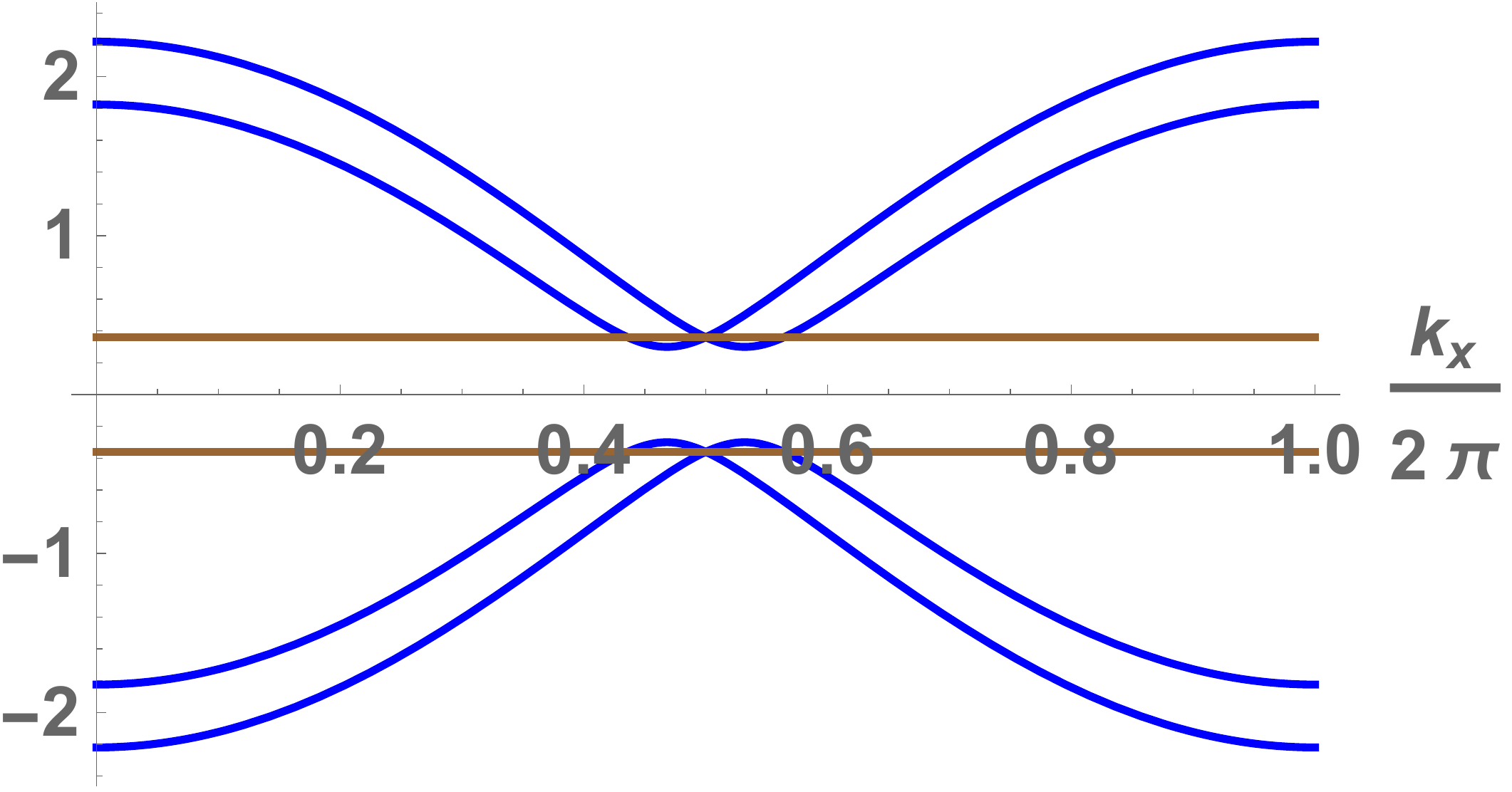} a)\\ } \end{minipage} \begin{minipage}[h]{.47\linewidth} \center{ \includegraphics[width=\linewidth]{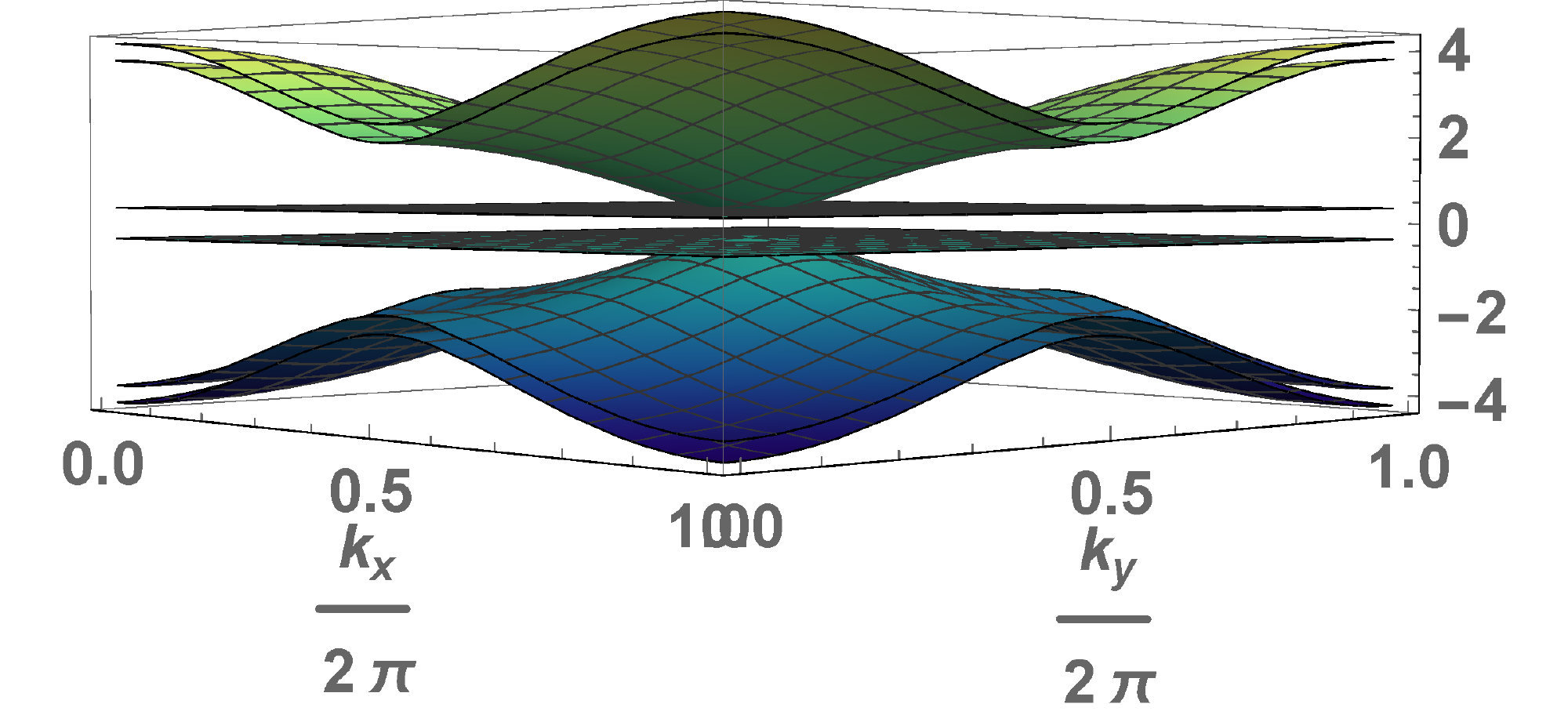} b)\\ } \end{minipage} 
\caption{(Color online) 
The spectrum of quasi-particle excitations of electron liquid  ( the gap  in the spectrum of charge excitations is equal to $2\lambda$) in the chain a) and square lattice b) as a function of the wave vector, calculated for $\lambda=0.3$, $h=0.2$  ($K=0.524$ for  chain and $K=0.597$ for square lqttice). } \label{fig:2} \end{figure}

\begin{figure}[tp] \centering{\leavevmode} \begin{minipage}[h]{.4\linewidth} \center{ \includegraphics[width=\linewidth]{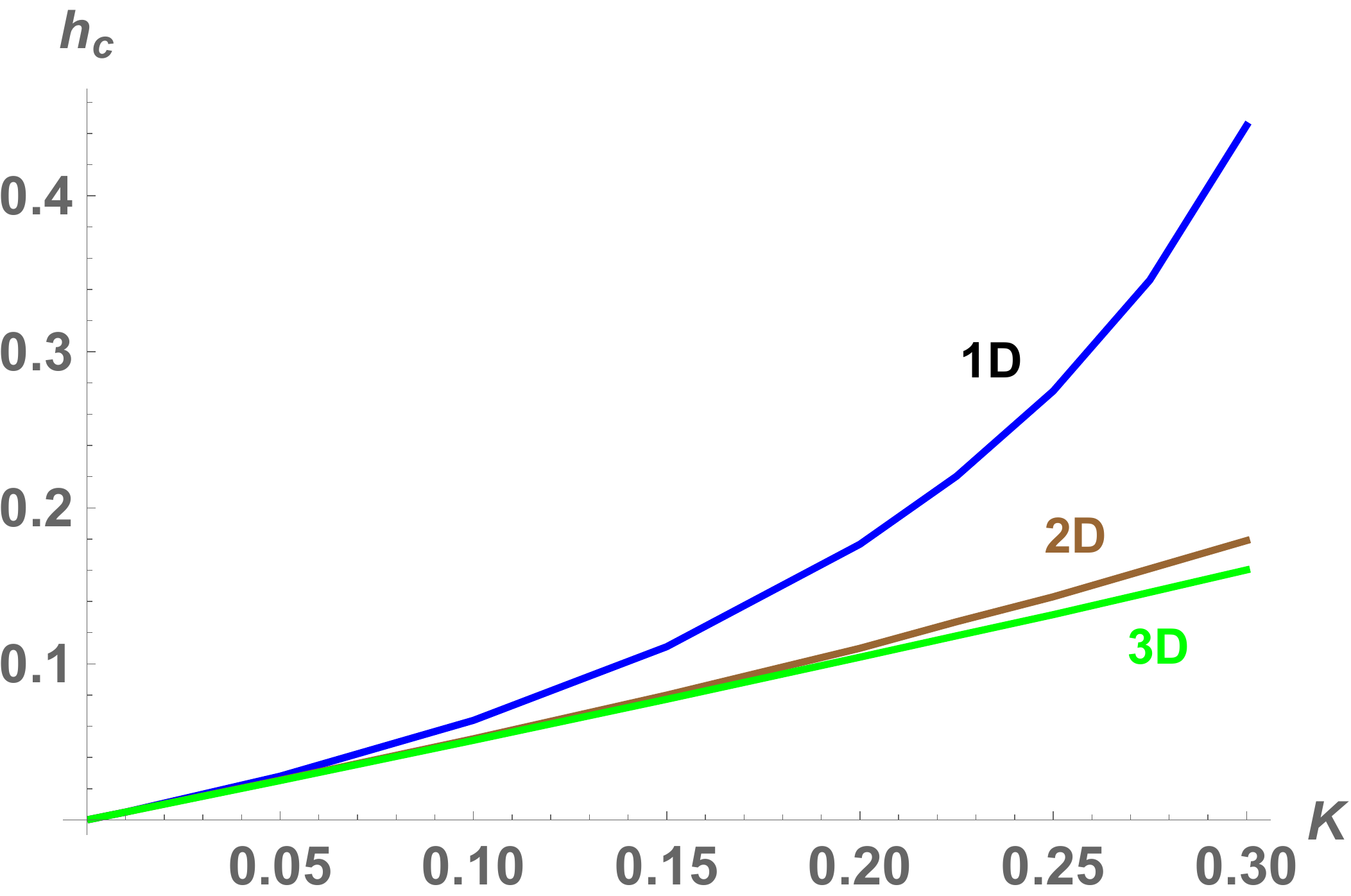}  } \end{minipage} \caption{(Color online) 
Critical value of magnetic field $h_c$, at which the gap in the quasi-particle spectrum  closes, as a function of the exchange integral, calculated for different dimension of the Kondo lattice. } \label{fig:3} \end{figure}

\begin{figure}[tp] \centering{\leavevmode} \begin{minipage}[h]{.45\linewidth} \center{ \includegraphics[width=\linewidth]{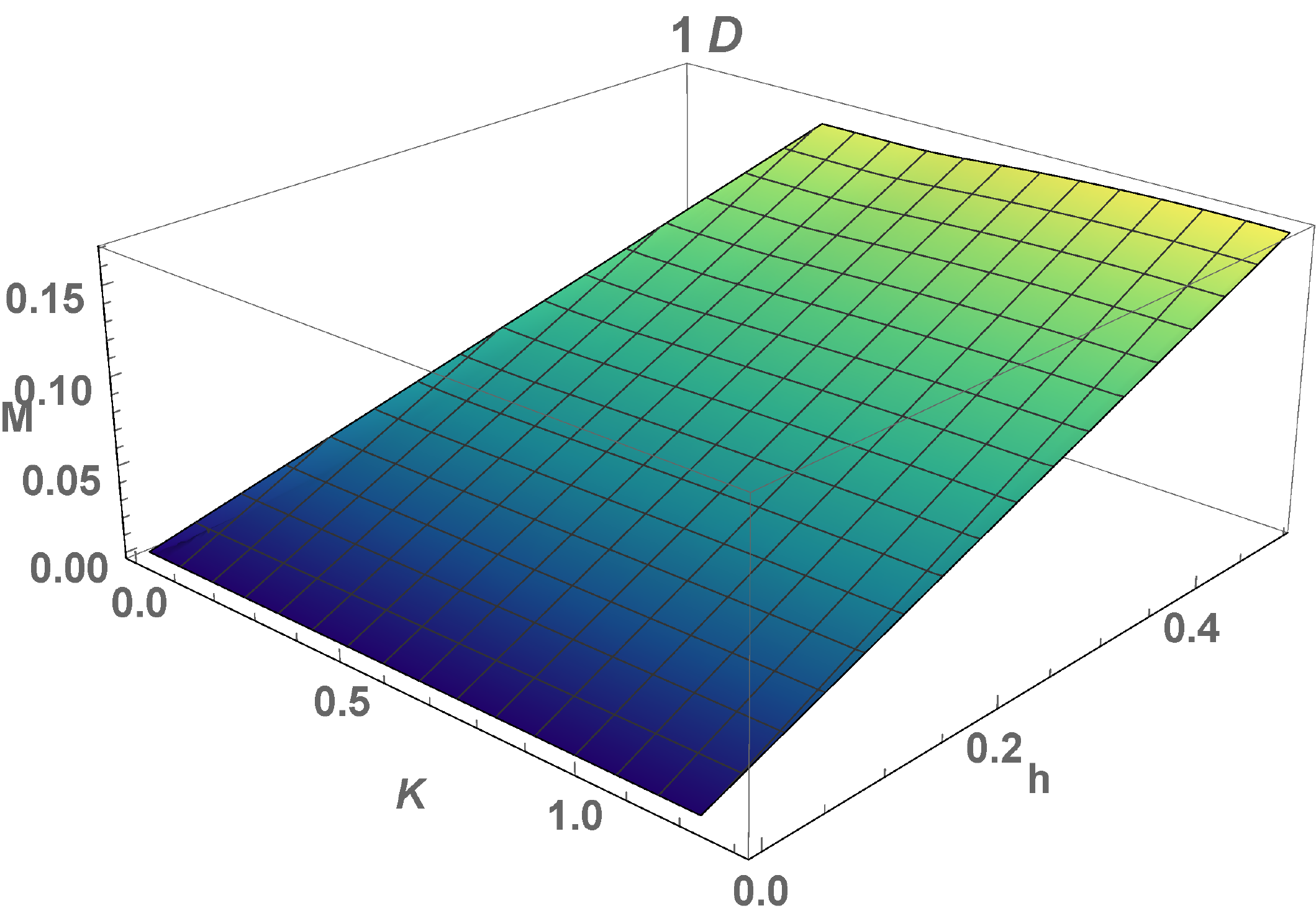} a)\\ 
} \end{minipage} \begin{minipage}[h]{.45\linewidth}\center{ \includegraphics[width=\linewidth]{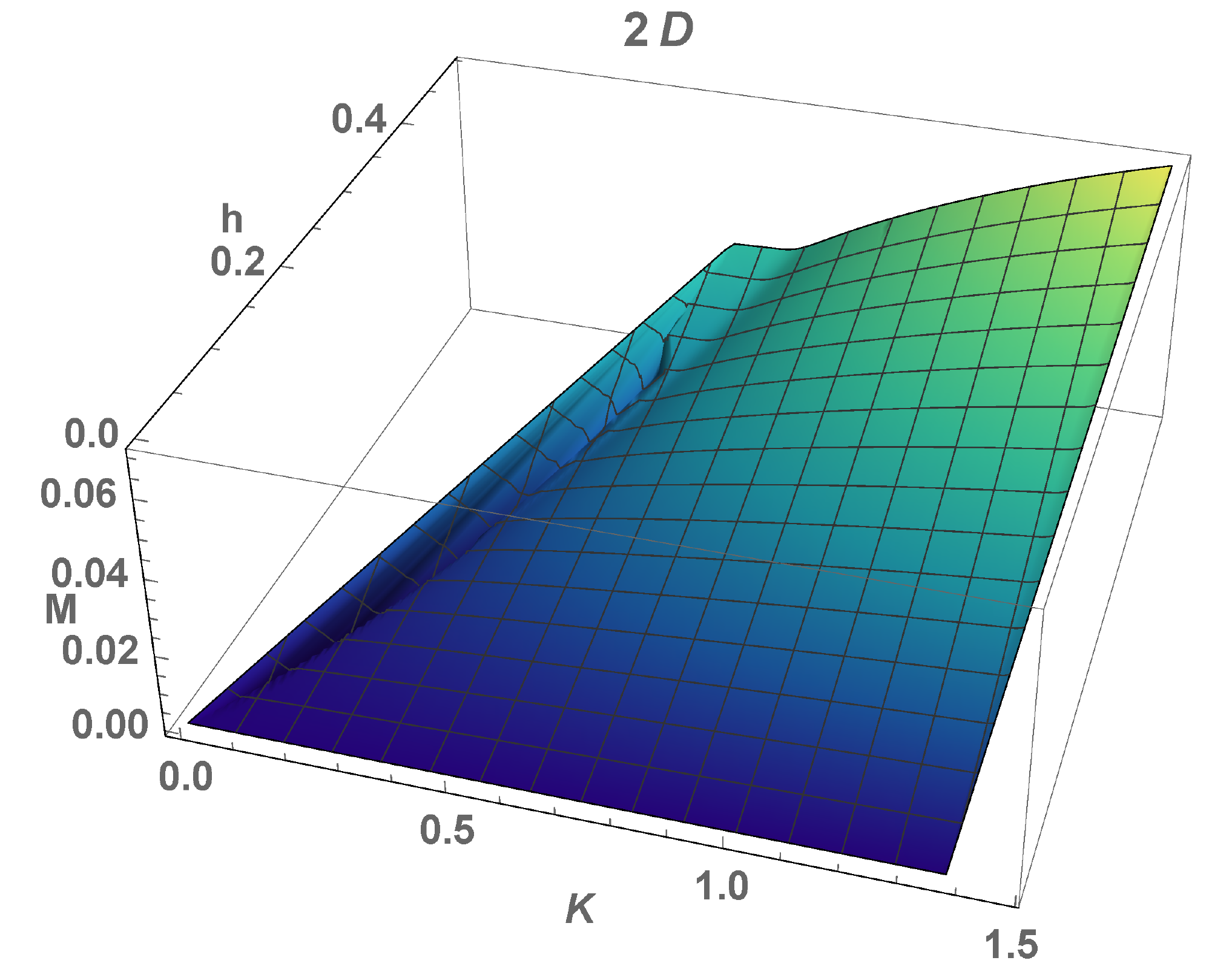} b)\\ 
} \end{minipage} \begin{minipage}[h]{.5\linewidth}
 \center{ \includegraphics[width=\linewidth]{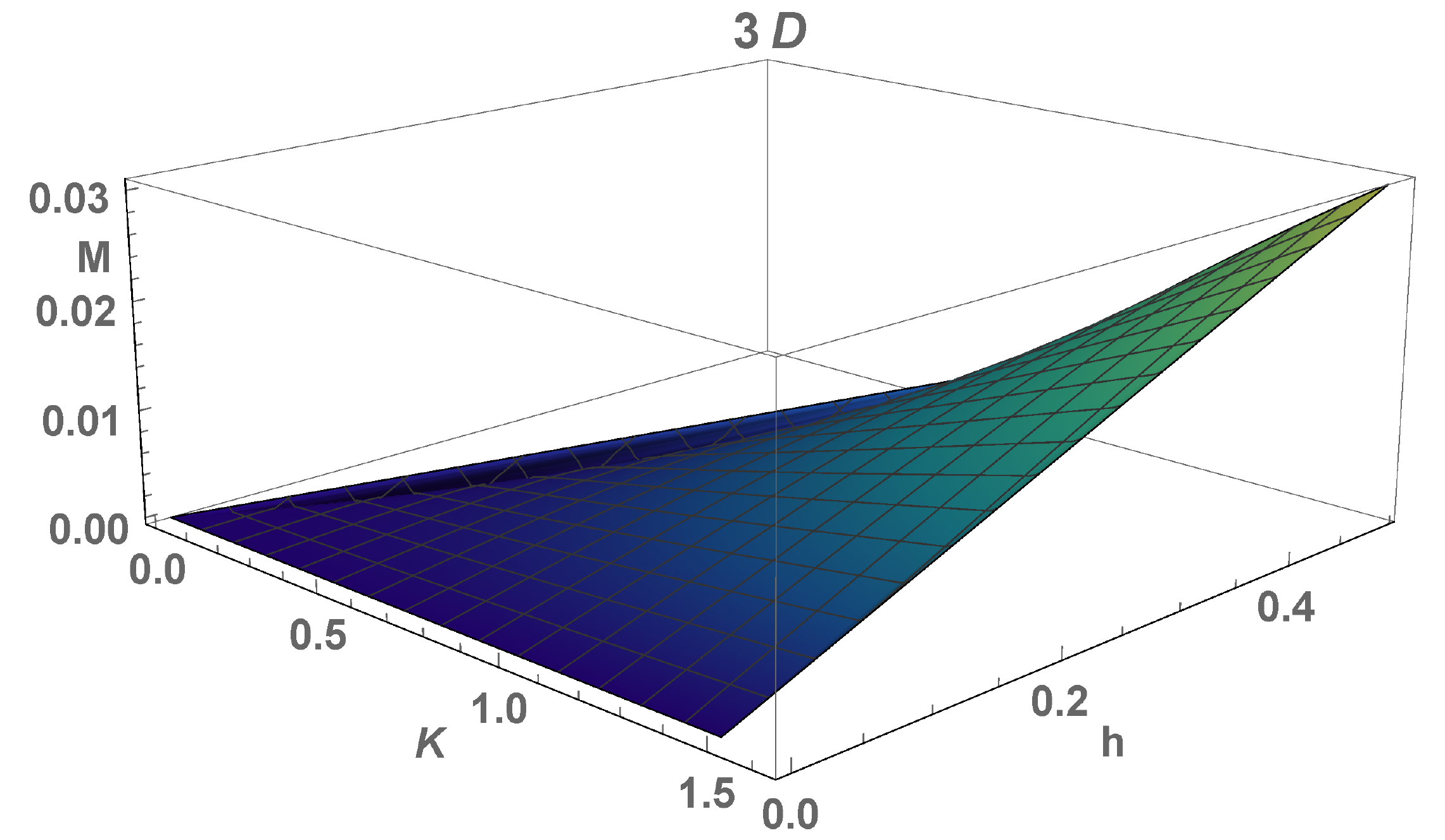} c)\\ 
} \end{minipage} \caption{(Color online)  Magnetization density  as a function of   the exchange  integral $K$ and magnetic field $h$ , calculated for the  chain a),  square b) and cubic c)  lattices . } \label{fig:4} \end{figure}

\begin{figure}[tp] \centering{\leavevmode} 
\begin{minipage}[h]{.4\linewidth} \center{ \includegraphics[width=\linewidth]{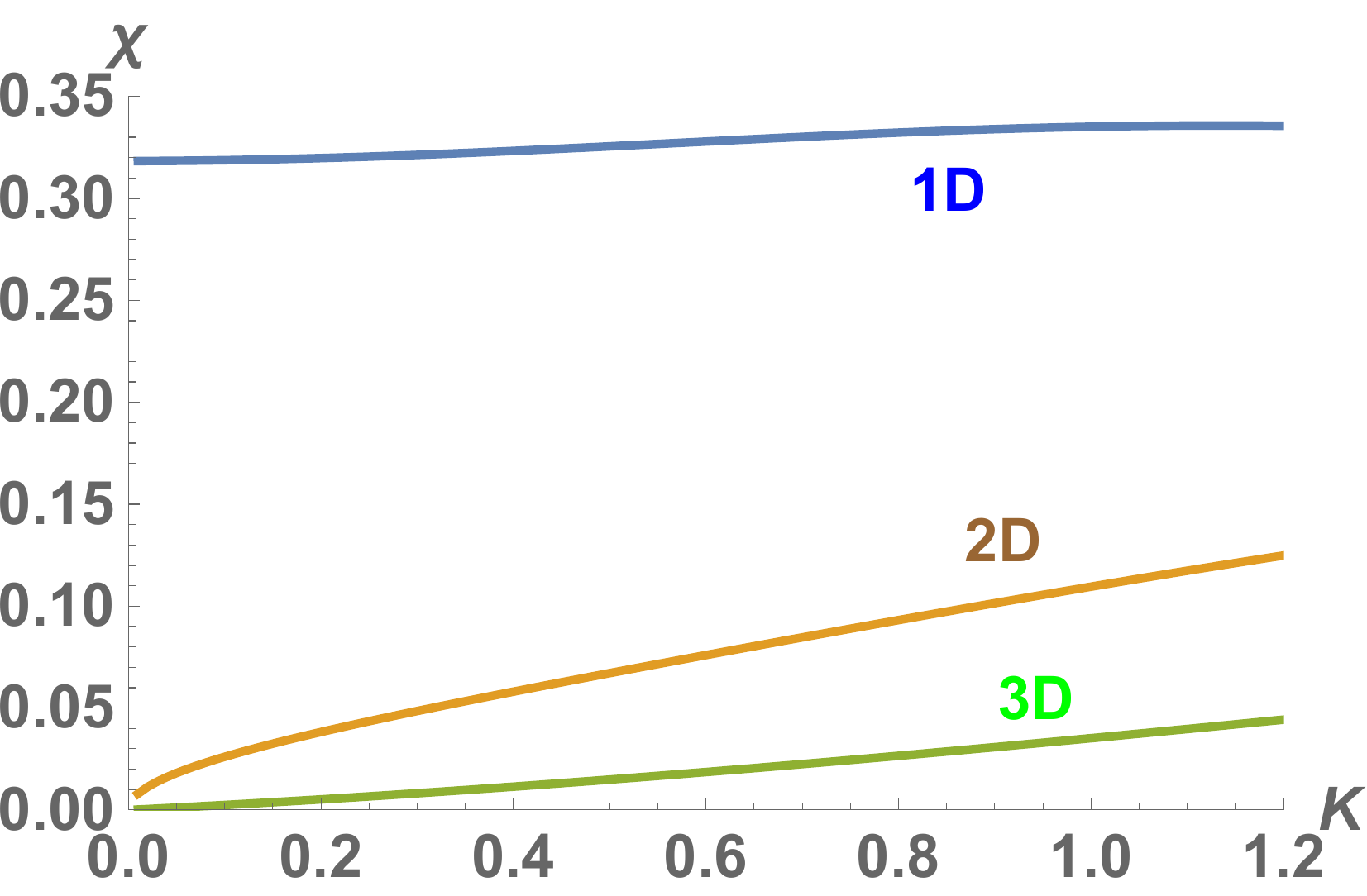}\\ } 
\end{minipage} 
\caption{(Color online) Static magnetic susseptibility as a function of an exchange integral calculeted for different dimension of the model. } \label{fig:5} \end{figure}

\begin{figure}[tp] \centering{\leavevmode} 
\begin{minipage}[h]{.4\linewidth} \center{ \includegraphics[width=\linewidth]{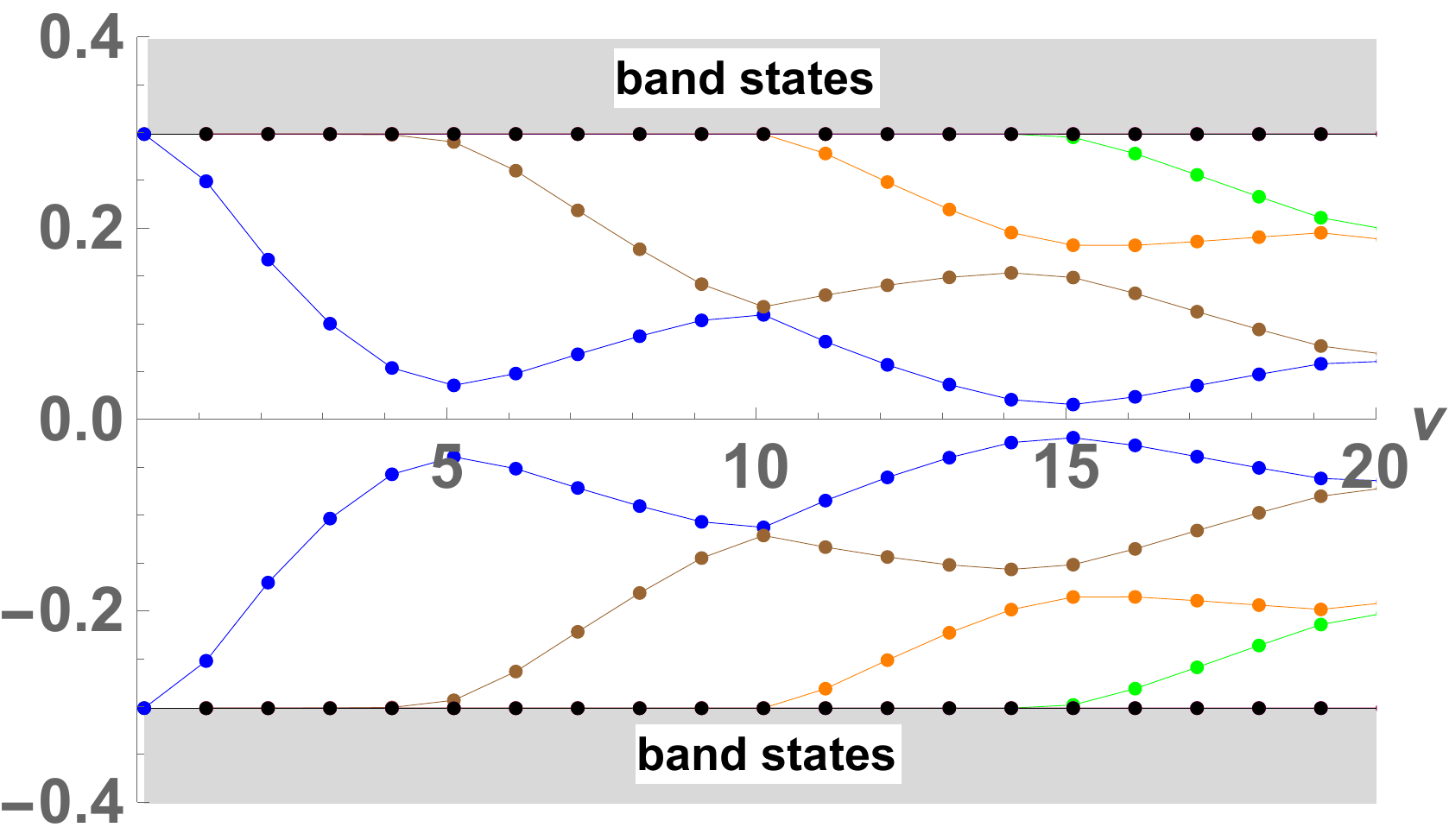}\\ } 
\end{minipage} 
\caption{(Color online) The excitation energies corresponding to the $Z_2$-field configuration with $\nu$ "defects" are calculated for various $\nu$ (marked with dots, the lines correspond to the excitation branches). Calculations were carried out for a chain with $\lambda=0.3$, $h=0.1$, $K=0.476$.
 } \label{fig:6} \end{figure}

\subsection*{Strongly anisotropic exchange interaction $J=0$, $K>0$}

For a strongly anisotropic exchange interaction, the $Z_2$- field is one-component, since $\Lambda=0$ and $\lambda \neq 0$. 
According to the numerical analysis, the solutions $\lambda_\textbf{j} =\lambda$ and $\lambda_\textbf{j+1}=-\lambda$   correspond to the minimum energy and for an arbitrary values of the exchange integral $K$ and magnetic field $h$.

In magnetic field the energies of the quasi-particle excitations transform to  $ \varepsilon_{s,+}(\textbf{k})=\pm\sqrt{\Lambda^2+\lambda^2+(h+|w(\textbf{k})|)^2}$,  
$ \varepsilon_{s,-}(\textbf{k})=\pm\sqrt{\Lambda^2+\lambda^2+(h-|w(\textbf{k})|)^2}$,  
 $\varepsilon_S=\pm\sqrt{\Lambda^2+\lambda^2+h^2}$. 
The spectrum is symmetrical with respect to zero energy or chemical potential, which is zero at half filling.

In the electron spectrum the gap opens at $\lambda\neq 0 $ and is equal to $2\lambda$ at $\Lambda =0$,  according to (4) its value is determined by $K$ and $h$. Using Eq (4) we numerically calculate $\lambda$ as function of $K$  and $h$ for the chain Fig1a),  square   Fig1b) and cubic  Fig1c) lattices.  Should be note an universal behavior of an electron liquid in KIS, the curves in Figs are similar for an arbitrary dimension. 
In a weak coupling limit at $K\to 0$  the last term  in Eq (4)  dominates, so  $\lambda \to \frac{K}{2}$.  We ilustrate the spectrum of the quasi-particle excitations in Fig2 a) for the chain and Fig2 b) for square lattice.  The magnetic field breaks the spin degeneracy of the spectrum of electrons, spreading the branches of the electron spectrum.
The $\lambda$- value (or the value of the gap) decreases with magnetic field. A critical value of the magnetic field $h_c$, at which the gap closes,  depends on $K-$value. Numerical calculations of $h_c$ are shown in Fig 3 (the curves are calculated for different dimension of the model). In magnetic  field $h_c$ the phase transition from insulator state to metal state is realized, in other words  KIS is stable at $h<h_c$.

Magnetic properties of an electron liquid in KIS are determined by both band electrons and local moments, 
they determine the uniform configuration of the $Z_2-$field. Electrons and local moments form singlet states  in the lattice with a double cell, which are not fixed in time. These singlet states (with different spins of electrons and local moments) are degenerated in energy in absence of magnetic field, so the magnetization density  $M=\frac{2}{N}\sum_{j}(s^z_j+S^z_j)$ is zero. The magnetic field does not break this degeration of energy for local moments, so the magnetization is detemined by an electron term $M=\frac{1}{N}\sum_{j}(m_{j,\uparrow}-m_{j,\downarrow})$:
 \begin{equation} 
M=\frac{h}{N}\sum_{\textbf{k}} \frac{1}{|\varepsilon_{s +}(\textbf{k})| +|\varepsilon_{s,-}(\textbf{k})|} (\frac{\lambda^2 + h^2- |W(\textbf{k})|^2 }{ |\varepsilon_{s,+}(\textbf{k})\varepsilon_{s,-}(\textbf{k})| } + 1  ).
\label{eq:H6} 
\end{equation}
The calculations  of the magnetization density $M$ as function of magnetic field and the exchange integral $K$ are presented in  Figs 4 for different dimension of the model.
Formula for a static magnetic susceptibility  leads from $M$   at $h\to 0$
$\chi =\frac{1}{N}\sum_{\textbf{k}} \frac{\lambda^2 }{(\lambda^2 +| W(\textbf{k})|)^{3/2}}$.
The value of a static magnetic susceptibility is calculated a function of the exchange integral for different dimension of the model, calculations are shown in Fig 5, the susseptibility is a monotonic fumction of $K$.

An uniform configuration stabilises the phase state with a double cell,  a  free configuration corresponds to gapless state with higher energy.  According to numerical calculations, a total energy of individual $\nu$ "defects" is greater than the energy of one "defect" of size $\nu$.  One "defect" with $\nu =1$ corresponds to a $Z_2$-field configuration with an incorrect value of $\lambda_j$, namely $-\lambda_j$ instead of $\lambda_j$, which occurs in a uniform configuration.
As an example, we present the calculations of the excitation energies in a chain with "defects" as a function of the defect size $\nu$.
Configurations with "defects" in an uniform  configuration have energies lying in the gap, a number of excitations increases with increasing 
$\nu$ , so that for $1<\nu<5$ only one excitation is split off from the continuous spectrum, for $5<\nu<10$ , $10<\nu<15$ there are 2 and 3 such states, respectively  (see in Fig 6).
We note, that the lattice with a double cell  is formed by an uniform configuration of  $\lambda $-field,  with no spin or charge density waves being realized.

\subsection*{Isotropic exchange interaction $J=K>0$}

As noted above, in the absence of a magnetic field  for an isotropic exchange interaction, the $\lambda$- and $\Lambda$-components of the $Z_2$-fields are equal and are solutions of Eq (3). Along with this solution, there are also a number of non-trivial solutions: $\lambda\neq 0$ and $\Lambda=0$ , $\lambda=0$ and $\Lambda \neq 0$. Three solutions of Eq (3) have the same energy of the ground state, which follows from numerical calculations of the ground state energy as funsction of $J$ at $h=0$ for different dimensions of the model. In the absence of an external magnetic field, the phase state of the electron liquid is degenerate.
 A magnetic field removes this degeneracy, namely, the solution $\lambda\neq 0$  and $\Lambda=0$  corresponds to a lower energy than solution 
$\lambda= 0$ and $\Lambda \neq 0$ for an arbitrary value of  magnetic field and an isotropic exchange integral. Another nontrivial  solution $\lambda \neq 0, \Lambda\neq 0$ not satisfy Eq(3) for arbitrary $h$. 
KIS is determined by the $XX$-exchange interaction (the value of $K$ in Hamiltonian (1)), the $ZZ$-exchange interaction (the value of $J$ in Hamiltonian (1)) does not participate in the formation of KIS. Scattering processes with spin flip lead to the formation of KIS in the Kondo lattice as it takes place in the Kondo problem.

In the case of an arbitrary spin $S$, Eq (4) for the amplitudes of the electron  wave function does not change. The energies of local moments change, so for $S=3/2$ $\varepsilon_S =\pm \frac{1}{2} (-1\pm\sqrt{5})\lambda$. As a result, the formula for the magnetization (5) does not depend on the magnetization of local moments, Eqs (3) are transformed taking into account the value of $\varepsilon_S$.

\section*{Conclusion}

We studied the behavior of electron liquid in the spin-$\frac{1}{2}$ Kondo lattice at half-filling for different dimension. 
Due to antiferromagnetic exchange interaction between band electrons  and local moments a static $Z_2$- field is formed. An uniform configuration of the $Z_2$ -field corresponds to ground state of electron liquid  in KIS, leads to formation of a lattice with a double cell. The spin or charge density waves are  not realized in  KIS.
In KIS the spectrum of quasi-particle excitations  is symmetric about zero energy, as it takes place for the Majorana spectrum.  
At  a critical value of a magnetic field, at which a gap closes, the phase transition to metal state is realized.

\section*{Acknowledgments} The author thanks the Weizmann Institute of Science and personally Prof. E. Berg for support.

 \section*{Author contributions statement} I.K. is an author of the manuscript 

\section*{Additional information} The author declares no competing financial interests. \\ 

\section*{Availability of Data and Materials} All data generated or analysed during this study are included in this published article.\\ Correspondence and requests for materials should be addressed to I.N.K. 
\end{document}